\documentstyle[ltwol]{article}

\def\be{\begin{equation}}
\def\ee{\end{equation}}
\def\bea{\begin{eqnarray}}
\def\eea{\end{eqnarray}}

\def\gsim{\begin{array}{c} > \\ \sim \end{array}}
\def\lsim{\begin{array}{c} < \\ \sim \end{array}}

\begin{document}

\title{CLOCKS, COMPUTERS, BLACK HOLES, SPACETIME FOAM,\\ AND HOLOGRAPHIC PRINCIPLE}

\author{Y. JACK NG}

\address{Department of Physics and Astronomy, University of North Carolina, 
Chapel Hill, NC 27599-3255, USA\\E-mail: yjng@physics.unc.edu}


\twocolumn[\maketitle\abstracts{ What do simple clocks, simple computers,
black holes, space-time foam, and holographic principle have in common?  I 
will show that the physics behind them is inter-related, linking together 
our concepts of information, gravity, and quantum uncertainty.  Thus, 
the physics 
that sets the limits to computation
and clock precision also yields
Hawking radiation of black holes and the 
holographic principle.  Moreover, the latter two strongly imply that 
space-time undergoes much larger quantum fluctuations than what the 
folklore suggests --- large enough to be detected with modern 
gravitational-wave interferometers through future refinements.}]

\section{Introduction \& Summary}
Computers and clocks are physical systems.  As such, they must obey the laws
of physics.  Here we show that, according to
the laws of quantum mechanics and gravitation,
the number $\nu$ of
operations per unit time, and the number $I$ of bits of information
in the memory space of a simple computer (``simple'' in the sense to be made 
precise below), are both limited 
such that their product is bounded by a universal constant given by $I
\nu^2 \lsim
t_P^{-2}$, where $t_P = (\hbar G / c^5)^{1/2} \sim 10^{-43} sec$ is the 
Planck time. 
We also show that the
total running time $T$ over which a simple clock can remain accurate, and the
smallest time interval $t$ that the clock is capable
of resolving, are bounded by $T t^{-3} \lsim t_P^{-2}$.  Remarkably, nature 
appears to make use of such simple clocks and simple computers, for
these bounds are saturated for black holes.  Then we show that the
physics that sets the limits~\cite{Ng} to computation and clock precision
is precisely the physics that
governs the quantum fluctuations of space-time leading to 
space-time foam~\cite{NvD1,Ka}.  These quantum 
fluctuations induce an uncertainty $\delta R$ in the measurement of distance
$R$ given by $R (\delta R)^{-3} \lsim l_P^{-2}$ (in contrast to 
$\delta R \gsim l_P$, independent of $R$, according to the 
forklore~\cite{MTW})
where $l_P = c t_P \sim 10^{-33} cm$ is the Planck length.  Interestingly, 
such
uncertainties in distance measurements are exactly what are necessary to 
yield the holographic 
principle~\cite{tH} which states that the number of degrees of freedom 
that can be put
into a region of space is bounded by the area of the region in Planck units.
And significantly, these quantum fluctuations of space-time 
\emph {can} be detected 
with modern gravitational-wave interferometers perhaps in the not-too-distant 
future.~\cite{AC,NvD2}

\section{Two Ingredients}
The ingredients we will use are the general principles of 
quantum mechanics and general relativity.  First let 
us follow Wigner~\cite{SW} to use quantum mechanics to set fundamental
limits on the mass $m$ of any clock.  For the clock to give time 
to within accuracy $t$ repeatedly 
throughout its running time $T$, it must have a spread in position 
$\delta R$ (throughout $T$) so small that the time at which 
a light quantum strikes it
(in order to read it) can be determined within accuracy $t$: 
$\delta R \lsim ct$.  But if the clock has a linear
spread of $\delta R$, then its momentum uncertainty is $\hbar (\delta
R)^{-1}$.  After a time $\tau$, its position spread grows to $\delta
R(\tau) = \delta R + \hbar \tau m^{-1} (\delta R)^{-1}$ with the minimum
at $\delta R = (\hbar \tau / m)^{1/2}$.
At the end of the total running time $T$, the linear spread can grow to
\begin{equation}
\delta R \gsim \left(\frac {\hbar T}{m}\right)^{1/2}.
\label{W1}
\end{equation}
But this spread in position is required to be less that $ct$.  
Hence, for a given $T$ and $t$, the
bound on $m$ reads
\begin{equation}
m \gsim \frac {\hbar}{c^2 t} \left(\frac {T}{t}\right).
\label{W2}
\end{equation}
This limit is \emph {more restrictive} than that 
given by Heisenberg's energy-time
uncertainty relation because the requirement that repeated measurement 
of time not
introduce significant inaccuracies over the total running time $T$ imposes
a more severe limit on its mass than a single simultaneous measurement 
of both the energy $mc^2$ and the time $t$.

The second ingredient comes from general relativity.  
Consider a simple clock 
consisting of two parallel mirrors (each of mass $m/2$)
between which bounces a beam of light.  On the one hand, for the clock to
be able to resolve time interval as small as $t$, the mirrors must be
separated by a distance $d$ with $d/c \lsim t$.  On the other hand,
$d$ is necessarily larger than the Schwarzschild radius $Gm/c^2$ of the
mirrors so that the clock does not collapse into a black hole.
From these two requirements, it follows~\cite{NvD1}
\begin{equation}
t \gsim \frac {Gm}{c^3}.
\label{vD1}
\end{equation}

\section{Simple Clocks}
For any simple clock with accuracy $t$ and total running time 
$T$, we can relate these two time scales by
substituting Eq. (\ref{W2}) into Eq. (\ref{vD1}) to get
\begin{equation}
T \lsim t \left( \frac {t}{t_P} \right)^2.
\label{Tt}
\end{equation}
Thus the more precise a clock is, i.e., the smaller t is, the shorter it
can keep accurate time, i.e., the smaller T is.  
But don't let any clock-salesman talk you into buying a relatively
inaccurate clock with the hope that it may last longer, for even
a femtosecond ($10^{-15}$ sec)
precision yields the bound $T \lsim 10^{34}$ years.  However, note that the 
bound on $T$ goes down rapidly as $t^3$.

Let us now explain what we mean by the qualification 
``simple'' characterizing the simple clock (and the simple computer).  
To do that, consider a large clock consisting of N identical small
clocks to keep time one after another. For large enough N, the T-t relation 
(Eq.~(\ref{Tt})) and Eq. (\ref{vD1}) are obviously violated for the 
large clock.  But fortunately, this argument is not valid if
we consider only those clocks for which no
such separation of components is involved.  They are what we call \emph {simple} clocks.  The same qualification will be understood to apply 
to \emph {simple} computers.  

\section{Simple Computers}
Since a computer can serve as a clock, the limit Eq.~(\ref{Tt}) also
applies to a computer.
To obtain a universal bound on the speed of computation and the 
memory space of any simple information processor,
we note that the fastest possible processing frequency is given by 
$\nu = t^{-1}$ and $T/t$, the maximum number
of steps of information processing, is, aside from factors like $ln 2$,
the amount of information $I$ that can be registered by the 
computer.  Then it is an easy matter 
to use the T-t relation 
in Eq.~(\ref{Tt}) to show that 
\begin{equation}
I \nu ^2 \lsim \frac {1}{t_P^2} \equiv \frac {c^5}{\hbar G} \sim 10^{86} /sec^2,
\label{N2}
\end{equation}
independent of the mass, size, and details of the simple computer. This
universal bound (valid for any \emph {simple} computer)
links together our concepts of information, gravity,
and quantum uncertainty.  We will see below that nature seems 
to respect this bound
which, in particular, is realized for black holes.  For
comparison, current laptops
perform about $10^{10}$ operations per sec on $10^{10}$ bits, yielding $I
\nu ^2 \sim 10^{30} / sec^2$, and the current largest $I \nu^2$ is around
$10^{39} / sec^2$ set by the largest IBM SP cluster.

\section{Black Holes}
If a black hole is used as a clock, then it is reasonable to expect
that the maximum running time of this gravitational clock and the minimum
time interval that it can be used to measure are given by 
the Hawking black hole life-time and the light travel time across 
the black hole's horizon respectively:
\begin{equation}
T \sim \frac {G^2 m^3}{\hbar c^4}, \hspace{0.3 in} t \sim \frac {Gm}{c^3}.
\label{B23}
\end{equation}
(Alternatively we can derive the above equation by 
appealing to Wigner's two inequalities Eq.~(\ref{W1}) 
and Eq. (\ref{W2}) and using the Schwarzschild radius of the black
hole as the minimum clock size.~\cite{Ba}  Thus, if we had not known of
black hole evaporation, this remarkable result would have implied 
that there is a maximum lifetime for a black hole when quantum 
observers are introduced.)  Now, according to the second half of 
Eq.~(\ref {B23}), the limit on $t$ as shown in Eq.~(\ref{vD1}) is saturated
for a black hole.  Furthermore, using both expressions in
Eq.~(\ref{B23}) one can easily
show that the bound given by Eq.~(\ref{W2}) is saturated.  It then
follows that the subsequent bounds (Eq.~(\ref{Tt}) and
Eq.~(\ref{N2})) are also saturated for black holes.  Thus, black hole are the
ultimate \emph {simple} clocks and the ultimate \emph {simple} computers.
It is curious
that although they can be very massive and large, black holes are basically
\emph {simple} --- a fact further supported by the no-hair theorem.

Here let us comment on the main difference between our approach and 
that of 
Lloyd~\cite{Ll} on the physical limits to computation.  
Lloyd's use of Heisenberg's energy-time
uncertainty principle to find $\nu$ is tantamount to putting $T \sim t$ in
Wigner's inequality (Eq.~(\ref {W2})).  Thus, while we have
introduced two time scales $T$ and $t$, Lloyd has introduced only
$t$.  But as the case of black holes shows, these two time scales are not the
same in general.  For a 1-kg black hole, according to Lloyd~\cite{Ll},
$\nu \sim 10^{51}/sec$ and $I \sim 10^{16}$ bits; but according to us,
Eq.~(\ref{W2}) and Eq.~(\ref{vD1}) yield
$\nu \sim 10^{35}/sec$ and $I \sim 10^{16}$ bits.  Thus we 
disagree with the limits given by Lloyd for the case of black holes.

\section{Space-time Foam}
To see the connection between the physical limits to computation and clock 
precison and the physics that governs the quantum
fluctuations of space-time (giving rise to space-time foam), 
let us consider measuring the
distance $R$ between two points.  We can put a clock at one of the points
and a mirror at the other point.  By sending a light signal from the clock
to the mirror in a timing experiment we can determine the distance.  But
the quantum uncertainty in the positions of the clock and the mirror
introduces an inaccuracy $\delta R$ in the distance measurement.  The same
argument used above to derive the T-t relation now yields a bound
for $\delta R$:
\begin{equation}
\delta R \left ( \frac {\delta R}{l_P}\right)^2 \gsim R,
\label{vD2}
\end{equation}
in a distance measurement.~\cite{NvD1}  In a time measurement, an analogous
bound is
given by Eq. (\ref{Tt}) with $T$ playing the role of the measured time and
$t$ the uncertainty.~\cite{NvD1}  This limitation to 
space-time measurements can be
interpreted as resulting from quantum fluctuations of space-time itself.
In other words, at short distance scales, space-time is foamy.  Thus the
same physics underlies both the foaminess of space-time and the limits to
computation and clock precision.  Not surprisingly, these bounds have
the same form.  

\section{Holographic Principle}
Furthermore, the same physics is behind the holographic 
principle~\cite {tH}.  To see this, consider a region of space 
with linear dimension R.  The
conventional wisdom claims that the region can be partitioned into cubes
as small as $(l_P)^3$.  It follows that the number of degrees of freedom
of the region is bounded by $(R/l_P)^3$, i.e., the volume of the region in
Planck units.  But the conventional wisdom is wrong this time, for according
to Eq. (\ref {vD2}), the smallest cubes into which we can partition the
region cannot have a linear dimension smaller than $(R l_P^2)^{1/3}$.
Therefore, the number of degrees of freedom of the region is bounded by
$[R/(R l_P^2)^{1/3}]^3$, i.e., the area of the region in Planck units,
as stipulated by the holographic principle.~\cite{NvD2} 

Two comments are in order: (1) Since black holes have an entropy given by
the event horizon in Planck units,~\cite{BeWh} the holographic bound is 
saturated for black holes.  We can see this also by using Eq. (\ref{B23}).
(2) As argued above, the holographic principle has its origin in
quantum fluctuations of space-time.  So, turning the argument around,
we believe the holographic principle alone suggests that
the quantum fluctuations of space-time are as given by Eq. (\ref{vD2})
and hence are much larger than what
the conventional wisdom\cite{MTW} leads us to believe.  

\section{Smoking Gun}
Both the Hawking radiation of black holes and the holographic principle
lend strong albeit indirect supports for the space-time foam model given
by Eq. (\ref{vD2}).  At present there is no solid direct evidence; after 
all, even on the size of the whole observable universe ($\sim 10^{10}$
ligh-years), Eq. (\ref{vD2}) yields a fluctuation of only about $10^{-13}$ 
cm.  Luckily, as pointed out recently~\cite{AC,NvD2}, modern 
gravitational-wave interferometers, through future refinements, may reach
displacement noise level low enough to test this space-time foam model. 
To see this, in any distance measurement that involves 
a time interval $\tau$, we note that there is a
minute uncertainty $\delta R \sim (c \tau l_P^2)^{1/3}$.  This uncertainty 
manifests itself as a displacement noise (in addition to other sources of
noises) that infests the interferometers.  The Fourier transform of this noise gives the displacement amplitude spectral density of frequency $f$:
\begin{equation}
S(f) \sim f^{-5/6}(c l_P^2)^{1/3}.
\label{spectral}
\end{equation}
By comparing the spectral density with the existing observed noise level
of $3 \times 10^{-17} {\rm cm-Hz}^{-1/2}$ near 450 Hz, the lowest noise level 
reached by the Caltech 40-meter interferometer, we obtain the bound
$l_P \lsim 10^{-27}$ cm.  The ``advanced phase'' of LIGO is expected to
achieve a noise level low enough to probe $l_P$ down to $10^{-31} cm$,
only about two orders of magnitude from what we expect it to be!  (For 
comparison, if the conventional space-time foam model~\cite{MTW} 
($\delta R \gsim l_P$) is correct, even the advanced phase of LIGO can 
probe $l_P$ only down to $10^{-17}$ cm.)  Eagerly we wait for 
our interferometer
colleagues to improve the sensitivity of modern gravitational-wave 
interferometers by
another two orders of magnitude to provide us with the ``smoking gun'' 
evidence of space-time foam (Eq. (\ref{vD2})).  
Perhaps then, for the first time, we will catch a glimpse of the 
very fabric of space-time!

\section*{Acknowledgments}
This work was supported in part by the U.S. Department of Energy under
\#DE-FG05-85ER-40219.

\end{document}